\documentclass[%
 preprint,superscriptaddress,
showpacs,preprintnumbers,
 amsmath,amssymb,
 aps,
 prc,
]{revtex4-1}
\usepackage{graphicx}
\usepackage{dcolumn}
\usepackage{bm}

\usepackage{subfigure}
\usepackage{mathrsfs}
\usepackage{multirow}
\begin{document}


\title{Brueckner-Hartree-Fock and its renormalized calculations for finite nuclei}

\author{B. S. Hu (ºú°Øɽ)}
\affiliation{State Key Laboratory of
Nuclear Physics and Technology, School of Physics, Peking University, Beijing 100871,
China}
\author{F. R. Xu (Ðí¸¦ÈÙ)}\thanks{frxu@pku.edu.cn}
\affiliation{State Key Laboratory of
Nuclear Physics and Technology, School of Physics, Peking University, Beijing 100871,
China}
\author{Q. Wu (ÎâÇ¿)}
\affiliation{State Key Laboratory of
Nuclear Physics and Technology, School of Physics, Peking University, Beijing 100871,
China}
\author{Y. Z. Ma (ÂíԶ׿}
\affiliation{State Key Laboratory of
Nuclear Physics and Technology, School of Physics, Peking University, Beijing 100871,
China}
\author{Z. H. Sun (ËïÖкÆ)}
\affiliation{State Key Laboratory of
Nuclear Physics and Technology, School of Physics, Peking University, Beijing 100871,
China}


\date{\today}

\begin{abstract}
We have performed self-consistent Brueckner-Hartree-Fock (BHF) and its renormalized theory
to the structure calculations of finite nuclei.
The $G$-matrix is calculated within the BHF basis,
and the exact Pauli exclusion operator is determined by the BHF spectrum.
Self-consistent occupation probabilities are included in the renormalized Brueckner-Hartree-Fock (RBHF).
Various systematics and convergences are studies.
Good results are obtained for the ground-state energy and radius.
RBHF can give a more reasonable single-particle spectrum and radius.
We present a first benchmark calculation with other {\it ab initio} methods
using the same effective Hamiltonian.
We find that the BHF and RBHF results are in good agreement with other {\it ab initio} methods.
\end{abstract}

\maketitle
\section{\label{sec:introduction} Introduction}

\begin{figure}
\includegraphics[scale=1.00]{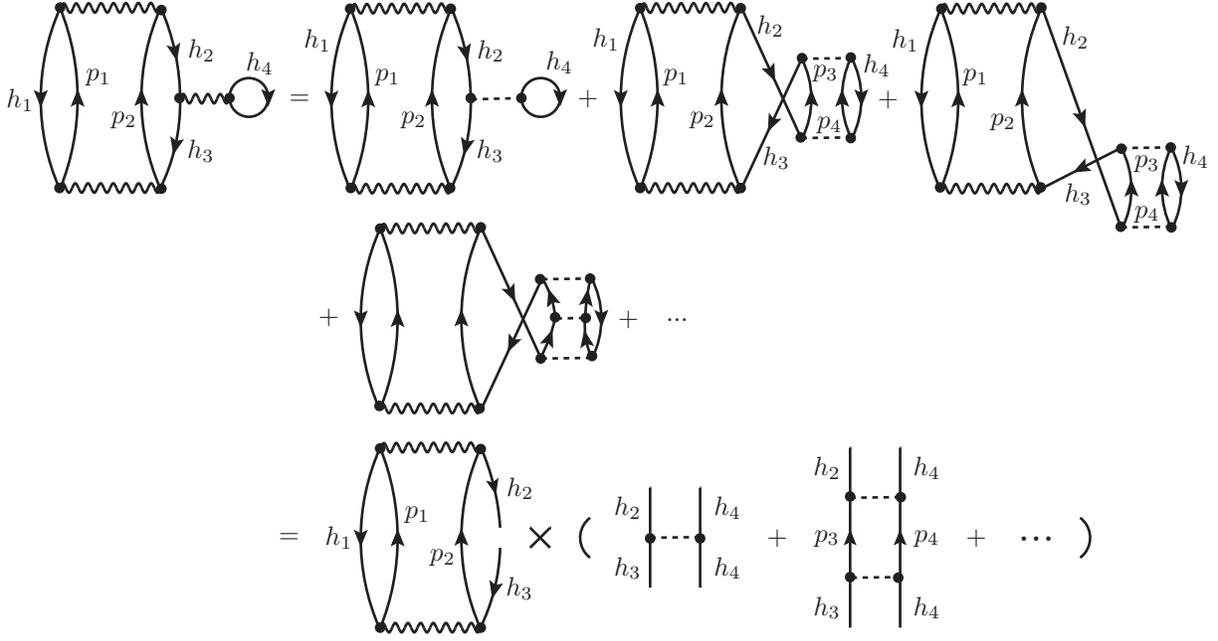}
\caption{\label{fig:on-shell} An example of the diagrams which can be included in hole-hole $G$-matrix bubble insertions
by putting $G$-matrix on the energy shell.}
\end{figure}

\begin{figure}
\includegraphics[scale=0.50]{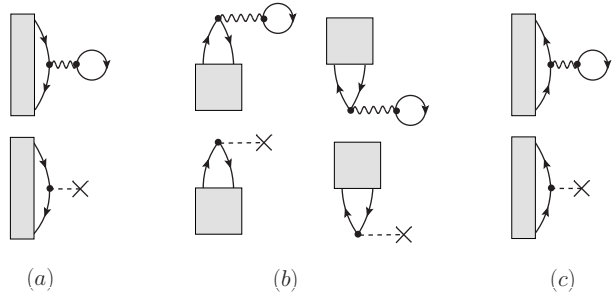}
\caption{\label{fig:diagrams-bhf}  Some diagrams summed in a BHF calculation.
Wavy lines signify $G$-matrix interactions while dashed 
line terminated by an $\times$ signifies negative single-particle potential $U$.}
\end{figure}

\begin{figure}
\includegraphics[scale=1.00]{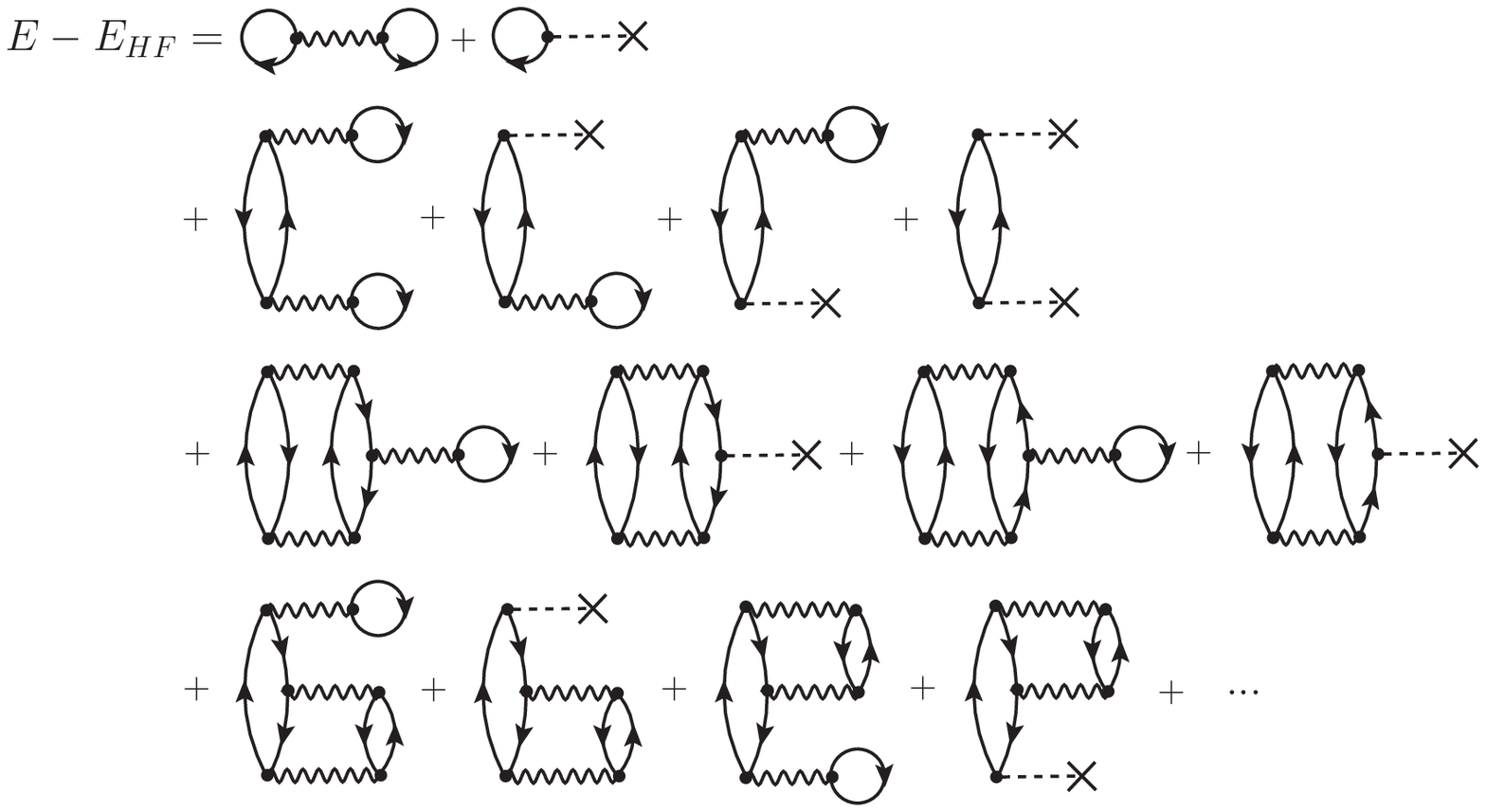}
\caption{\label{fig:e-cancel} First-, second-, and some third-order anti-symmetrized Goldstone diagrams of energy corrections 
in Brueckner-Goldstone expansion~\cite{nla1}.
Wavy lines signify $G$-matrix interactions while dashed 
line terminated by an $\times$ signifies negative single-particle potential $U$.}
\end{figure}

\begin{figure}
\includegraphics[scale=0.50]{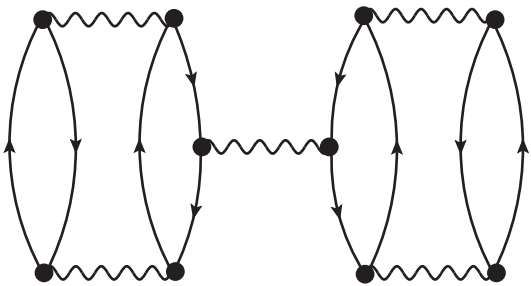}
\caption{\label{fig:rbhf}  Typical diagram cancelled in RBHF formalism by including occupation
probabilities in the definition of the single-particle potential $U$.}
\end{figure}

One of the fundamental goals in nuclear structure theory is to understand the properties of strongly interacting $A$-nucleon system
in terms of the realistic nucleon-nucleon ($NN$) interaction between the constituent, protons and neutrons.
Where the ``realistic $NN$ interaction'' means $NN$ potential that
provides high-quality descriptions of
the deuteron properties and
the $NN$ scattering phase shifts up to a certain energy, typically up to 350 MeV at laboratory energies.
Hartree-Fock (HF) method is one of the simplest approximations for solving the many-body quantum system,
which is based on a single Slater determinant of single-particle states.
These single-particle states are eigenstates of the one-body HF potential $U$,
which is determined from the two-body $NN$ interaction $V$
including the Coulomb interaction by a self-consistent calculation.
The conventional HF method describes the motion of nucleons in the average field of other nucleons
and neglects higher-order correlations.
Obviously, the HF approach is cannot to describe full correlations
when using realistic interactions.
Brueckner-Hartree-Fock (BHF) theory gives an improved definition of one-body potential $U$ by replacing $V$ to a so-called reaction matrix G ($G$-matrix),
which corresponds to a summation of ladder diagrams to infinite orders
and formally represents an effective two-body interaction allowing for many-body correlation effects.
In this theory, the important diagrams in perturbation expansion are summed by introducing the operator $G$-matrix,
and the residual effects of $V$ not allowed by $U$ can be small.
The important diagrams include not only the ladder diagrams to infinite orders,
but also some diagrams that can be included in hole-hole and particle-hole $G$-matrix bubble insertions
by putting $G$-matrix on the energy shell, e.g., Fig.~\ref{fig:on-shell},
or in particle-particle bubble insertions by the off-shell prescription.
These hole-hole and particle-hole bubble insertions can be exactly cancelled by choosing $U$ \cite{nla1}.
It means that the lower panel of Fig.~\ref{fig:diagrams-bhf} exactly cancel the upper panel.
These particle-particle bubble insertions cancel the
three-body cluster diagrams as much as possible by the off-shell prescription~\cite{RevModPhys.39.745,PhysRev.177.1519,nla1}.
Figure~\ref{fig:e-cancel} gives the Brueckner-Goldstone expansion for ground-state energy, 
where $V$ is replaced by $G$ in the perturbation expansion~\cite{bartlett2009}, and ladder diagrams are omitted.
We can see that the bubble insertions can be cancelled by choosing $U$ \cite{nla1}.
The Renormalized Brueckner-Hartree-Fock (RBHF) approach \cite{PhysRevC.4.81,PhysRevLett.24.400,PhysRevC.9.1221,PhysRevC.92.034312} 
is a slight modification of the BHF field,
which takes into account the depletions of
the normally occupied single-particle states
resulting from many-body correlations
and cancels a large class of additional diagrams
(called saturation-potential diagrams, or rearrangement diagrams),
e.g., shown in Fig.~\ref{fig:rbhf},
in calculating the ground-state energy and single-particle energies.
It has pointed out that the gap between occupied and unoccupied states is decreased,
the ground-state energy is increased, and the radius means square is decreased, 
comparing the RBHF to BHF \cite{PhysRevC.4.81,PhysRevC.1.1644,PhysRevC.10.2080,PhysRevC.92.034312}.
But in this work we will give a different conclusion.

Standard realistic interactions, such as
CD-Bonn \cite{PhysRevC.63.024001},
Nijmegen \cite{PhysRevC.49.2950},
Argonne $\upsilon_{18}$ \cite{PhysRevC.51.38},
INOY \cite{PhysRevC.69.054001}, and
chiral potential \cite{PhysRevC.68.041001,Machleidt20111},
exhibit strong short-range correlations
which cause convergence problem in the calculations of nuclear structure.
This problem is evident for these potentials that have a so-called hard-core \cite{cnp2}.
The matrix elements of such a potential, $\langle \phi(r)|V_{NN}|\phi(r) \rangle$,
in an uncorrelated two-body wave function $\phi(r)$ will become very large or even diverge,
since the uncorrelated wave function is different from zero also for
relative distance $r$ smaller than the hard-core radius.
Alternatively, realistic interactions can also be expressed in momentum space.
Then strong repulsive core as well as tensor force of the potential are directly associated with
the coupling between low-momentum and high-momentum parts of the potential matrix elements.
This implies that the basis expansion is significantly complicated for solving the many-body Schr\"{o}dinger equation.
For example, in a harmonic-oscillator (HO) basis, which is the most common choice for the finite nuclei calculation,
convergence is substantially slowed by the need to accommodate these strong short-range correlations.
So none of them can be used as ``bare'' in nuclear structure calculations without renormalization
or a large-enough truncated HO basis.
To deal with the strong short-range correlations and speed up the convergence,
realistic forces are usually processed by certain renormalizations.
A traditional approach is the G-matrix renormalization in the Brueckner-Bethe-Goldstone theory
\cite{PhysRev.97.1353,goldstone1957,PhysRev.129.225}.
Recently, a new class of renormalization methods has been developed,
including $V_{\text{low-}k}$ \cite{PhysRevC.65.051301,Bogner20031},
similarity renormalization group (SRG) \cite{PhysRevC.75.061001},
Okubo-Lee-Suzuki \cite{okubo01111954,Suzuki01121980,Suzuki01071982,Suzuki01081983,Suzuki01121982,Suzuki01121994}
and unitary correlation operator method (UCOM) \cite{PhysRevC.72.034002,Roth201050}.

As the above statement, we know that the $G$-matrix in traditional BHF theory plays two roles.
One is the ingredient of constructing mean field $U$ to beyond HF
by including the important high-order perturbation terms.
The other is to deal with the strong short-range correlations.
In BHF theory,
the $G$-matrix is defined by the Bethe-Goldstone equation,
  \begin{eqnarray}
  G(\omega)=\hat{V} + \hat{V}\dfrac{Q}{e} G(\omega),
  \label{g}
  \end{eqnarray}
where the energy denominator $e=\omega - \hat{H}_{0}(1)-\hat{H}_{0}(2)+ i\eta$,
$\omega$ is the starting energy, $\hat{H}_{0}(i=1,2)$ is the single-particle Hamiltonian,
and the Pauli exclusion operator $Q$ forbids
any components with two of the interacting nucleons scattered into these states occupied by other nucleons.
In self-consistent BHF theory,
the Pauli exclusion operator $Q$ is determined by the BHF spectrum, namely,
expressing the Eq.~(\ref{g}) in the self-consistent BHF basis.
In practical calculation of the $G$-matrix in the past, there are two fundamental equations at least,
  \begin{eqnarray}
  G_{0}=\hat{V}_{NN} + \hat{V}_{NN}\dfrac{Q_{0}}{e_{0}} G_{0}
  \end{eqnarray}
and the BBP identity \cite{PhysRev.129.225},
  \begin{eqnarray}
  G=G_{0}^{\dag}+G_{0}^{\dag} \left(\dfrac{Q}{e}-\dfrac{Q_{0}}{e_{0}} \right) G,
  \label{bbp}
  \end{eqnarray}
where the subscript zero implies that $G_{0}$ is a reference $G$-matrix obtained
by using approximate Pauli operator $Q_{0}$ and energy denominator $e_{0}$ relative to the ``true'' reaction matrix $G$.
Calculating $G_{0}$ in relative and center-of-mass coordinates is a key step for BHF to deal with the short-range correlations.
There are many approximate $Q_{0}$ in the history of BHF development.
One example is the so-called angle-averaged Pauli operator approximation \cite{Sauer1970467,cnp2}.
Another example is Eden and Emery approximation \cite{eeprsa1958} that the Pauli operator is diagonal in the center-of-mass representation.
When we neglect the Pauli operator, namely, $Q_{0}=1$, this is the reference-spectrum method.
After getting $G_{0}$, the Eq.~(\ref{bbp}) is used to get the ``true'' $G$.
The most of methods do not use a self-consistent Pauli exclusion operator in the calculation of $G$-matrix.
They corrected $G_{0}$ in a HO representation, implying that the distinction between occupied and
unoccupied states in Pauli exclusion operator is determined
by a HO spectrum rather than from the self-consistent BHF spectrum.
If less approximation will be used, 
we need to exactly solve the Bethe-Goldstone equation, i.e., Eq.~(\ref{g}), in the self-consistent BHF basis, 
as in this work will do.

The main purpose of this paper is to calculate the bulk properties of
doubly, closed-shell nuclei using self-consistent BHF and RBHF approaches with realistic nuclear force.
The main differences comparing to the traditional BHF calculations are that
we use $V_{\text{low-}k}$ method rather than $G$-matrix to deal with the strong short-range correlations,
and the Bethe-Goldstone equation, Eq.~(\ref{g}), is exactly solved in the self-consistent BHF basis.
We also first time compare the calculations of BHF and RBHF with other \emph{ab initio} methods
using the same effective Hamiltonian.
This paper is organized as follows.
In Section~\ref{sec:1}, we present the formalisms of BHF and RBHF approaches for finite nuclei.
In Section~\ref{sec:2}, results of the calculation and benchmarks with other \emph{ab initio} methods are summarized.
A summary and outlook is given in Section~\ref{sec:summary}.

\section{\label{sec:1}Theoretical framework}

\subsection{ The effective Hamiltonian }

The intrinsic Hamiltonian of the $A$-nucleon system used in this work reads
\begin{eqnarray}
\text{\^{H}} &&=
\displaystyle\sum_{i=1}^{A} \left(1-\dfrac{1}{A}\right) \frac{\vec{p}_{i}^{2}}{2m} +
\displaystyle\sum_{i<j=1}^{A}   \left(\hat{V}^{(2)}_{ij}-\frac{\vec{p}_{i}\cdot\vec{p}_{j} }{mA} \right)
\nonumber \\ &&
= \displaystyle\sum_{i=1}^{A}\hat{T_{i}}+\displaystyle\sum_{i<j=1}^{A}\hat{V}_{ij},
\label{eq1}
\end{eqnarray}
where $\hat{V}^{(2)}=\hat{V}_{NN}+\hat{V}_{\text{coul.}}$, $\hat{V}_{NN}$ is the $NN$ interaction,
and $\hat{V}_{\text{coul.}}$ is the coulomb interaction.
We do not include a three-body interaction.
In the present work, the $\hat{V}_{NN}$ is derived from Argonne $\upsilon_{18}$ potential \cite{PhysRevC.51.38}
by the $V_{\text{low-}k}$ technique.
The $V_{\text{low-}k}$ method is a renormalization group approach
and is used to soften the short-range repulsion and short-range tensor components of initial interaction.
It integrates out the high-momentum components of $V_{NN}$ in momentum space
while preserves two-nucleon observables for relative momenta up to the cutoff $\Lambda$.
This process leads to high- and low-momentum parts of Hamiltonian being decoupled,
which means the renormalized potential becomes softer and more perturbative than initial potential.
The Lee-Suzuki projection method is used in our calculations to obtain the low momentum Hamiltonian \cite{Bogner20031,PhysRevC.73.064307}.

\subsection{Brueckner-Hartree-Fock}

In this paper, we use the letters $h_{1}$, $h_{2}$, ... to indicate the occupied
levels (hole states) in HF (or BHF) states, the letters $p_{1}$, $p_{2}$, ...
to the empty levels (particle states), and the letters $a$, $b$, ... to any states (either hole or particle).
As shown in Section~\ref{sec:introduction}, the BHF approach has almost the same formalism with HF approach
except that the HF single-particle potential $U$ is redefined by $G$-matrix.
A conventional choice for the matrix elements of BHF potential $U$ is
\begin{eqnarray}
\label{bhf potential}
\langle a|U|b \rangle=
\left\{
\begin{array}{lll}
\dfrac{1}{2} \displaystyle\sum_{h \leq \varepsilon_{F}}
\langle ah| G(\varepsilon_{a}+\varepsilon_{h})+G(\varepsilon_{b}+\varepsilon_{h})
|bh\rangle \qquad \text{for} \; a,b\leq \varepsilon_{F}\\
\displaystyle\sum_{h \leq \varepsilon_{F}}\langle ah| G(\varepsilon_{a}+\varepsilon_{h}))
|bh\rangle  \qquad \; \quad \quad \qquad \qquad \text{for} \; a  \leq \varepsilon_{F}, b > \varepsilon_{F}\\
\displaystyle\sum_{h \leq \varepsilon_{F}}\langle ah| G(\varepsilon_{b}+\varepsilon_{h}))
|bh\rangle  \qquad \; \quad \quad \qquad \qquad \text{for} \; a  > \varepsilon_{F}, b \leq \varepsilon_{F}\\
\dfrac{1}{2} \displaystyle\sum_{h \leq \varepsilon_{F}}\langle ah| G(\bar{\varepsilon}_{a}+\varepsilon_{h})+G(\bar{\varepsilon_{b}}+\varepsilon_{h})
|bh\rangle \qquad \text{for} \; a,b > \varepsilon_{F}
\end{array} \right. ,
\end{eqnarray}
where $\varepsilon_{F}$ is the Fermi energy,
$\bar{\varepsilon}_{a}=2\varepsilon_{0}-\varepsilon_{a}$, and
$\varepsilon_{0}$ is the average energy of occupied single-particle states.
For the elements of $U$ involving hole states, i.e., $\langle h |U|a\rangle$,
the on-energy-shell definition of $G(\omega)$ yields an exact cancellation of
hole-hole and particle-hole diagrams with bubble insertions
by application of the Bethe, Brandow and Petschek (BBP) theorem \cite{PhysRev.129.225}.
Here ``on-energy-shell'' means $\omega$ is equal either to
the energy of the initial two-particle state for $\hat{H}_{0}(1)+\hat{H}_{0}(2)$
 or to the energy of the final two-particle state in Eq.~(\ref{g}).
In all other cases, we say the $G(\omega)$ is calculated by off energy shell.
The definition of $\omega$
for particle-particle elements, $\langle p_{1}|U|p_{2} \rangle$,
is a somewhat controversial matter
for the corresponding particle-bubble diagrams require an off-energy-shell calculation.
Since the $\langle p_{1}|U|p_{2} \rangle$ depends on the excitation energy of the remainder diagram,
a self-consistent treatment of particle-bubble  diagrams is quite complicated.
It has been found that
the total contribution of
summing all the three-body cluster diagrams to the ground-state energy of nuclear matter is very much
smaller than the contribution of the single-particle bubble diagram in literature~\cite{RevModPhys.39.745}.
Thus if one requires the elements of $U$ between particle states to cancel the
three-body cluster diagrams they will not have to be very large.
Some works set  $\langle p_{1}|U|p_{2} \rangle$ to be zero.
We choose another prescription in Refs.~\cite{PhysRev.177.1519,nla1} as shown in Eq.~(\ref{bhf potential}).
We will discuss the details using the two different prescriptions of the  $\langle p_{1}|U|p_{2} \rangle$.

At present, the BHF calculations are limited to the spherical, closed-shell nuclei.
The spherical symmetry preserves
the quantum numbers of the orbital momentum ($l$), the total angular momentum ($j$)
and its projection ($m_j$) for the BHF single-particle states.
In the spherical closed shell,
the BHF single-particle eigenvalues are independent of the magnetic quantum number $m_j$,
which leads to a $2j+1$ degeneracy.
So we will calculate the elements of $G(\omega)$ in the angular momentum coupled scheme.
The BHF states are denoted by $|a\rangle =|\nu l j m_{t}\rangle$ with
$\nu$ and $m_{t}$ for other quantum numbers and isospin projection, respectively.
We define an anti-symmetrized two-particle state coupled to good angular momentum $J$ and projection $M$,
\begin{eqnarray}
| (ab)J M\rangle=\dfrac{1}{\sqrt{(1+\delta_{ab}) }}\displaystyle\sum_{m_{a},m_{b}}
\langle j_{a} m_{a}j_{b} m_{b}| J M \rangle
| (a m_{a})(b m_{b})\rangle.
\end{eqnarray}
To define the single-particle potential $U$, we need the elements of $G(\omega)$.
In self-consistent BHF,
$G$-matrix, i.e., Eq.~(\ref{g}), must satisfy that
the single-particle Hamiltonian $\hat{H}_{0}=\hat{T}+\hat{U}$,
and the Pauli operator $Q$ is determined by the BHF spectrum.
However, the elements of $G(\omega)$ within BHF basis,
i.e., $\langle (ab)JM|G(\omega)|(cd)JM \rangle$,
does not be completely defined, because $G(\omega)$ also depends on the  starting energy $\omega$.
In this work, we expand the on-energy-shell or off-energy-shell choice of $\omega$
as above discussion to the all elements of $G(\omega)$.
Then the Eq.~(\ref{g}) reads
\begin{eqnarray}
\langle (ab)JM|G(\varepsilon^{'}_{a}+\varepsilon^{'}_{b})|(cd)JM\rangle=&& \langle (ab)JM|\hat{V}|(cd)JM\rangle
\nonumber \\ &&
+ \dfrac{1}{2} \displaystyle\sum_{r,s > \varepsilon_{F}} \langle (ab)JM|\hat{V}|(rs)JM\rangle
\nonumber \\ && \times
\dfrac{ 1+\delta_{rs} }
{\varepsilon^{'}_{a}+\varepsilon^{'}_{b}-\varepsilon_{r}-\varepsilon_{s}}
\langle (rs)JM|G(\varepsilon^{'}_{a}+\varepsilon^{'}_{b})|(cd)JM\rangle
\label{g-matrix}
\end{eqnarray}
with
\begin{eqnarray}
\varepsilon^{'}_{b}=
\left\{
\begin{array}{c}
\varepsilon_{b} \qquad \qquad \; \text{for} \; b \leq \varepsilon_{F}  \\
2\varepsilon_{0}- \varepsilon_{b} \qquad  \text{for} \; b > \varepsilon_{F}
\end{array} \right..
\end{eqnarray}
We solve Eq.~(\ref{g-matrix}) by matrix inversion method.

After defining $U$, we obtain the BHF equations,
\begin{eqnarray}
\langle a|\hat{H}_{0}| b\rangle=\langle a|(\hat{T}+U)| b\rangle
=\varepsilon_{a}\delta_{ab}.
\label{bhf_eqs}
\end{eqnarray}
Solved these equations by iteration \cite{ring1980nuclear},
we can get the BHF single-particle energies $\varepsilon_{a}$ and states $|a \rangle$.
This process is similar to the solution of Spherical HF as shown in Ref.~\cite{PhysRevC.94.014303}.
Then the bulk properties of finite nuclei can be gotten by HF framework.
For example, the ground-state energy is given by,
\begin{eqnarray}
\begin{array}{ll}
  E_{\text{BHF}}=\displaystyle\sum_{h_{1}=1}^{A} \langle h_{1}|\hat{T}|h_{1} \rangle
  +\dfrac{1}{2} \displaystyle\sum_{h_{1},h_{2}=1}^{A}
  \langle h_{1}h_{2} |G(\varepsilon_{h_{1}}+\varepsilon_{h_{2}})|h_{1}h_{2}\rangle
  \\ \quad \quad \;\; =\displaystyle\sum_{h_{1}=1}^{A} \varepsilon_{h_{1}}
  -\dfrac{1}{2} \displaystyle\sum_{h_{1},h_{2}=1}^{A}
  \langle h_{1}h_{2} |G(\varepsilon_{h_{1}}+\varepsilon_{h_{2}})|h_{1}h_{2}\rangle.
  \end{array}
  \label{e_bhf}
  \end{eqnarray}

\subsection{Renormalized Brueckner-Hartree-Fock}

The RBHF approach is a slight modification of the BHF field,
which takes into account the depletions of the normally occupied single-particle states resulting from two-body correlations.
It includes occupation-probability diagrams (also called saturation-potential diagrams, or rearrangement diagrams)
compared with BHF
in calculating the ground-state energy and single-particle energies.
In RBHF, the occupation probability is defined by
\begin{eqnarray}
P_{h_{1}}=\left[ 1- \displaystyle\sum_{h_{2}}
\langle h_{1}h_{2}| \frac{\partial G(\omega)}{ \partial \omega }
\Bigg|_{\omega=\epsilon_{h_{1}}+\epsilon_{h_{2}}}
|h_{1} h_{2}\rangle P_{h_{2}} \right]^{-1},
\label{p}
\end{eqnarray}
where the renormalized single-particle energy is
\begin{eqnarray}
\epsilon_{h_{1}}=\langle h_{1}|T|h_{1} \rangle +
 \displaystyle\sum_{h_{2}}
\langle h_{1}h_{2}|G(\omega=\epsilon_{h_{1}}+\epsilon_{h_{2}})|h_{1} h_{2}\rangle P_{h_{2}}.
\end{eqnarray}
From Eqs.~(\ref{g}) and~(\ref{p}), we can easily get \cite{PhysRevC.1.1644},
\begin{eqnarray}
\frac{\partial G(\omega)}{ \partial \omega }
=-G(\omega) \left( \dfrac{Q}{\omega-H_{0}(1)-H_{0}(2)} \right)^{2} G(\omega),
\end{eqnarray}

\begin{eqnarray}
P_{h_{1}}=\left[ 1+ \displaystyle\sum_{h_{2}p_{1}p_{2}}
\dfrac{1}{2}
\left( \dfrac{\langle h_{1}h_{2}|G(\omega=\epsilon_{h_{1}}+\epsilon_{h_{2}}) |p_{1} p_{2}\rangle }
{\epsilon_{h_{1}}+\epsilon_{h_{2}}-\epsilon_{p_{1}}-\epsilon_{p_{2}} } \right)^{2}
 P_{h_{2}} \right]^{-1}.
 \label{p1}
\end{eqnarray}
We use the self-consistent iteration procedure to solve the above occupation probability Eq.~(\ref{p1}).

The ground-state energy in RBHF theory is
\begin{eqnarray}
E_{\text{RBHF}}=&&\displaystyle\sum_{h_{1}}\langle h_{1}|T|h_{1} \rangle +
\frac{1}{2} \displaystyle\sum_{h_{1},h_{2}}
\langle h_{1}h_{2}|G(\epsilon_{h_{1}}+\epsilon_{h_{2}})|h_{1} h_{2}\rangle
P_{h_{1}}P_{h_{2}}
\nonumber \\ &&
+\displaystyle\sum_{h_{1},h_{2}} (1-P_{h_{1}})
\langle h_{1}h_{2}|G(\epsilon_{h_{1}}+\epsilon_{h_{2}})|h_{1} h_{2}\rangle
P_{h_{2}}.
\label{energy}
\end{eqnarray}
The last term is an ``over-counting correction"
which must be included in the total energy
when single-particle energies are renormalized with occupation probabilities \cite{PhysRevC.1.1644,PhysRev.152.863,RevModPhys.39.771}.

\section{\label{sec:2}Calculations and discussions}

In this section, we apply the BHF and RBHF methods outlined in Section~\ref{sec:1} to
closed-shell nuclei ($^{4}$He, $^{16}$O and $^{40}$Ca).
The $V_{\text{low-}k}$ effective interaction derived from Argonne $\upsilon_{18}$ potential \cite{PhysRevC.51.38}
is adopted to benchmark against other  \emph{ab initio} calculations.
A sharp cutoff $\Lambda=1.9$ fm$^{-1}$ is used for the $^{4}$He calculations
to compare with the Faddeev-Yakubovsky (FY) \cite{PhysRevC.70.061002,PhysRevC.76.044305} and CC results.
For $^{16}$O and $^{40}$Ca, we take  $\Lambda=2.1$ fm$^{-1}$
to compare with the results of the CC \cite{PhysRevC.76.044305} and importance-truncated NCSM (IT-NCSM) \cite{PhysRevLett.99.092501}.
In our calculations, the self-consistent BHF (or RBHF) basis is expanded by the HO basis.
The HO basis is truncated by a cutoff according to
the number $N_{\rm shell} = \text{max}(2n+l +1)$,
where the labels are standard with $n$ and $l$
for the radial and orbital angular momentum quantum numbers of the HO basis,
respectively.
 $N_{\rm shell}$ indicates how many major HO shells are included in the truncation.

\begin{figure}
\includegraphics[scale=0.43]{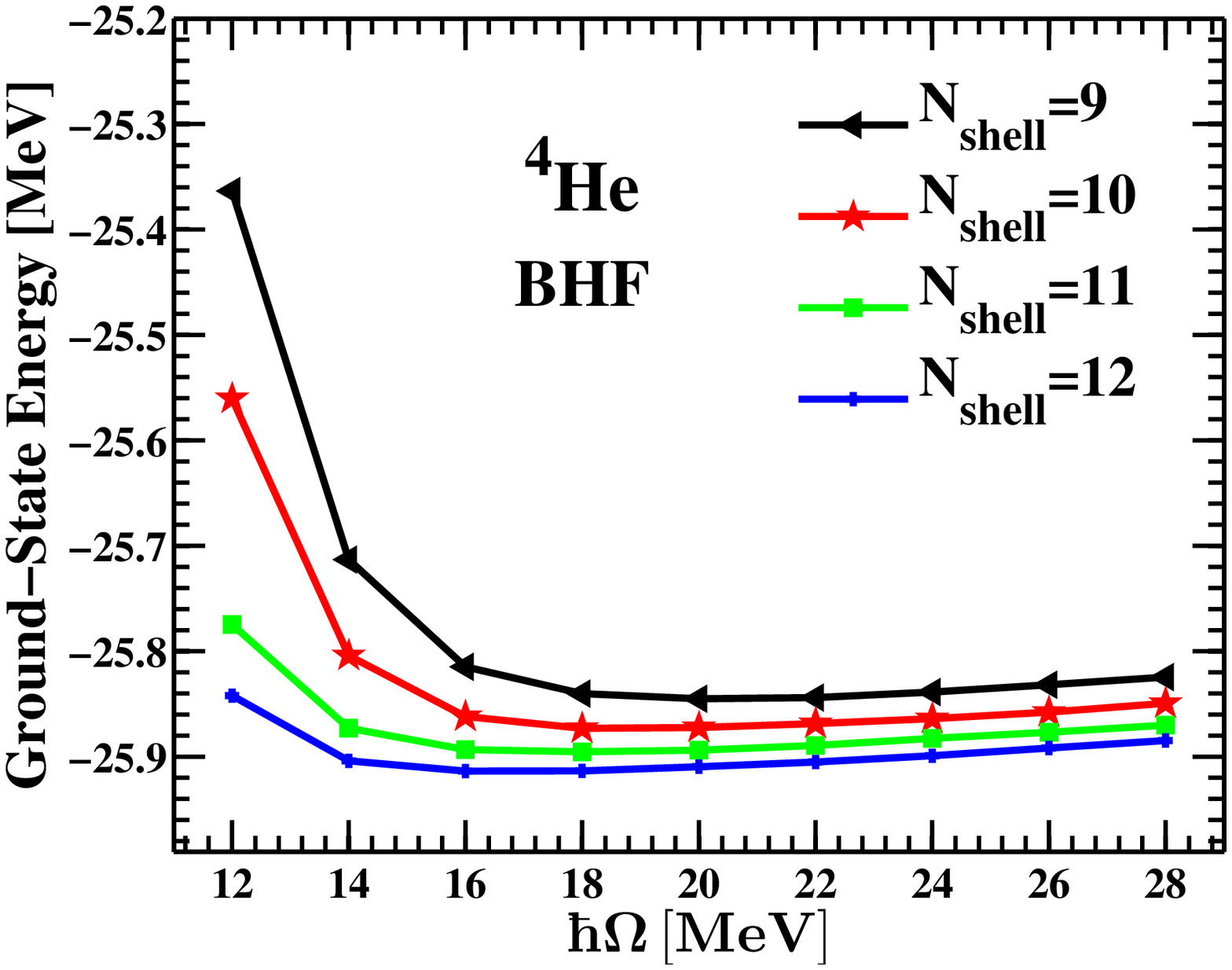}
\includegraphics[scale=0.43]{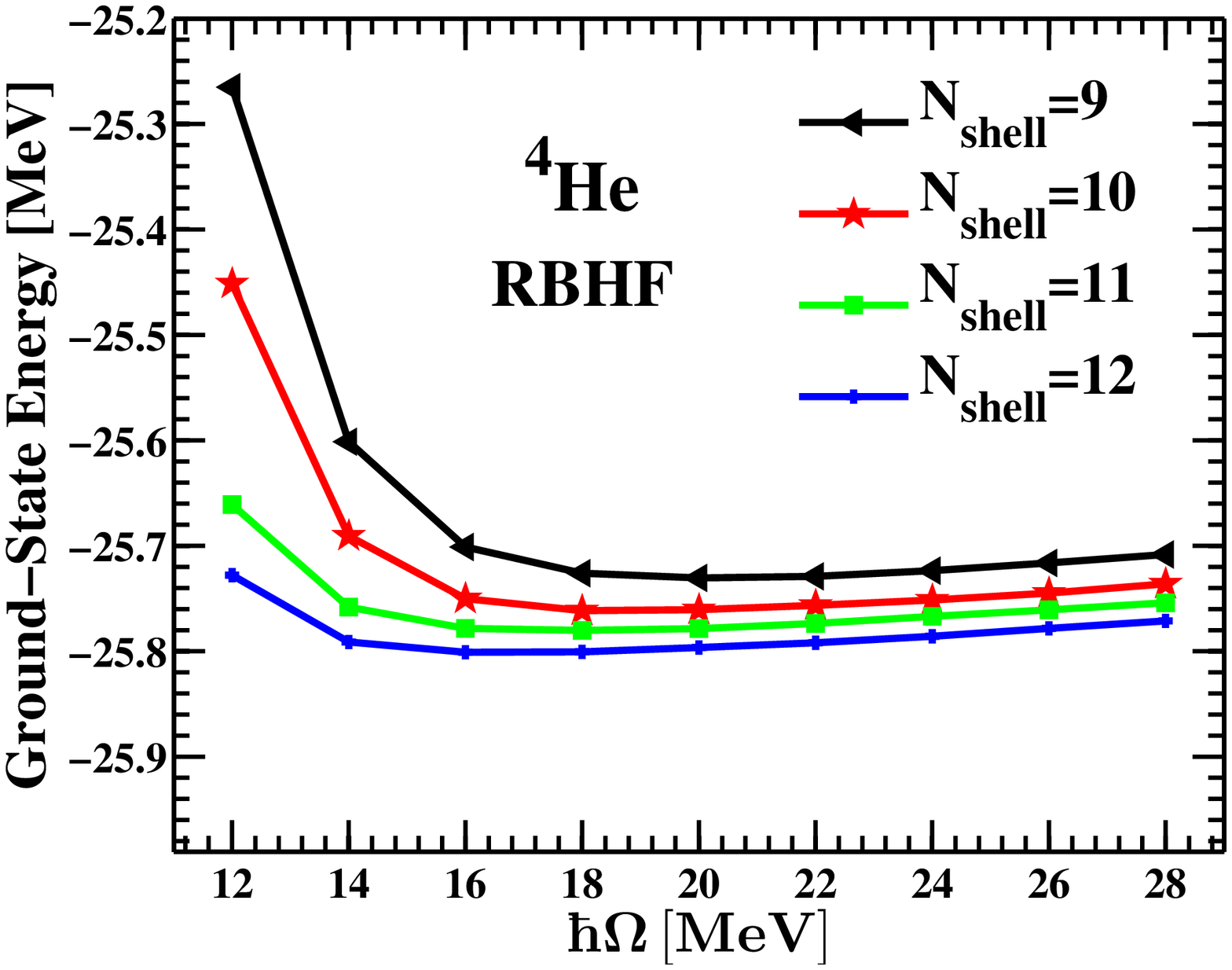}
\\
\includegraphics[scale=0.43]{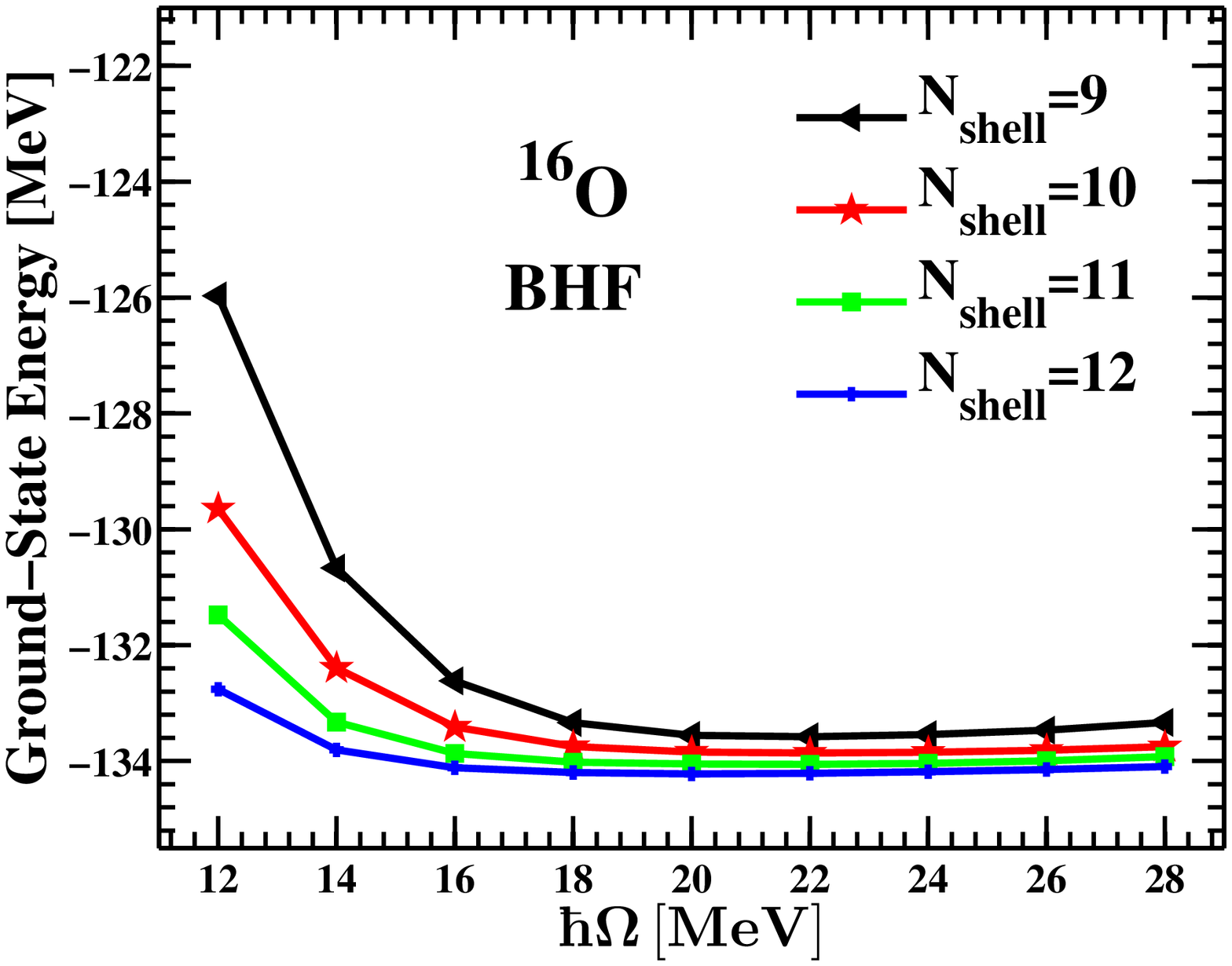}
\includegraphics[scale=0.43]{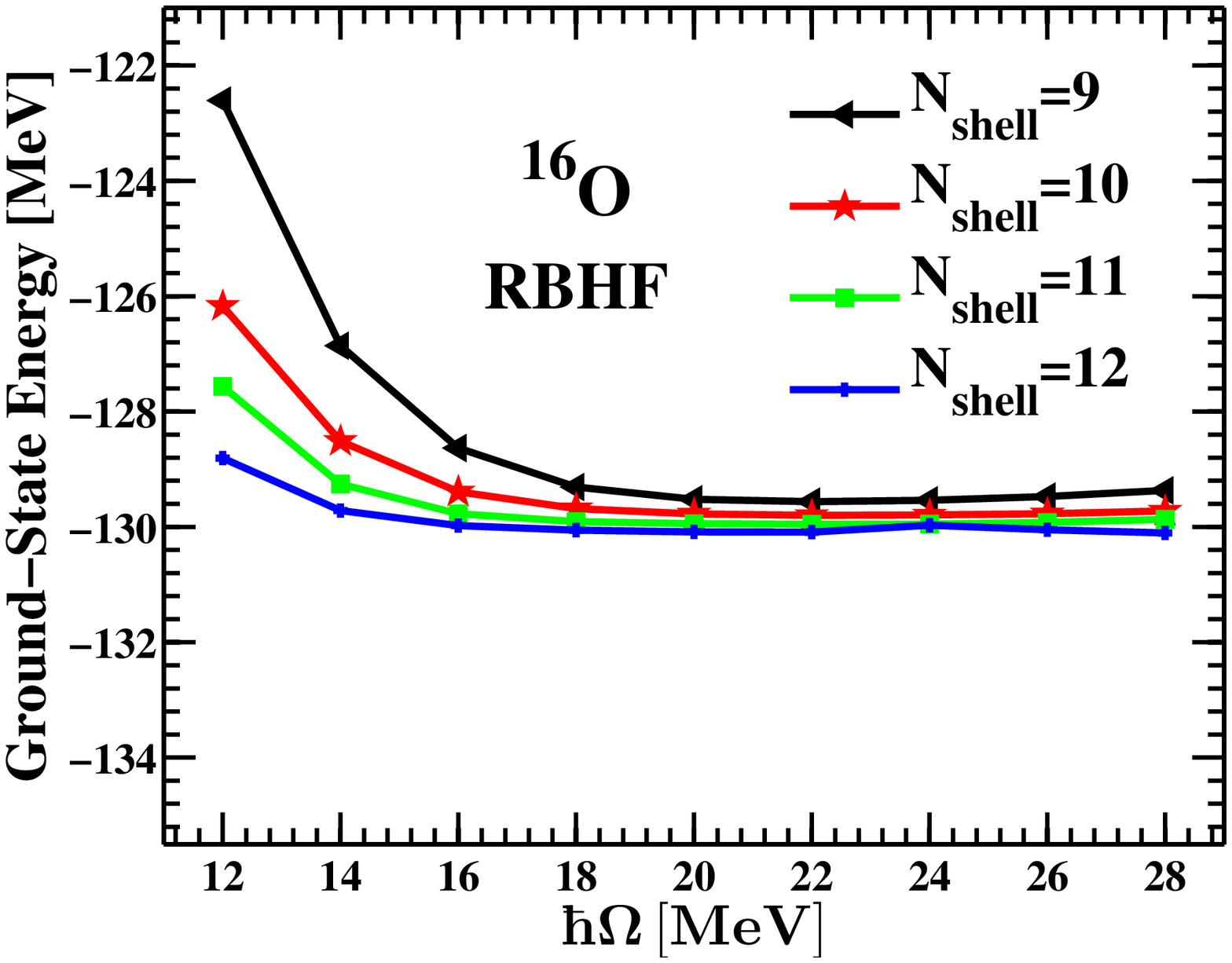}
\\
\includegraphics[scale=0.43]{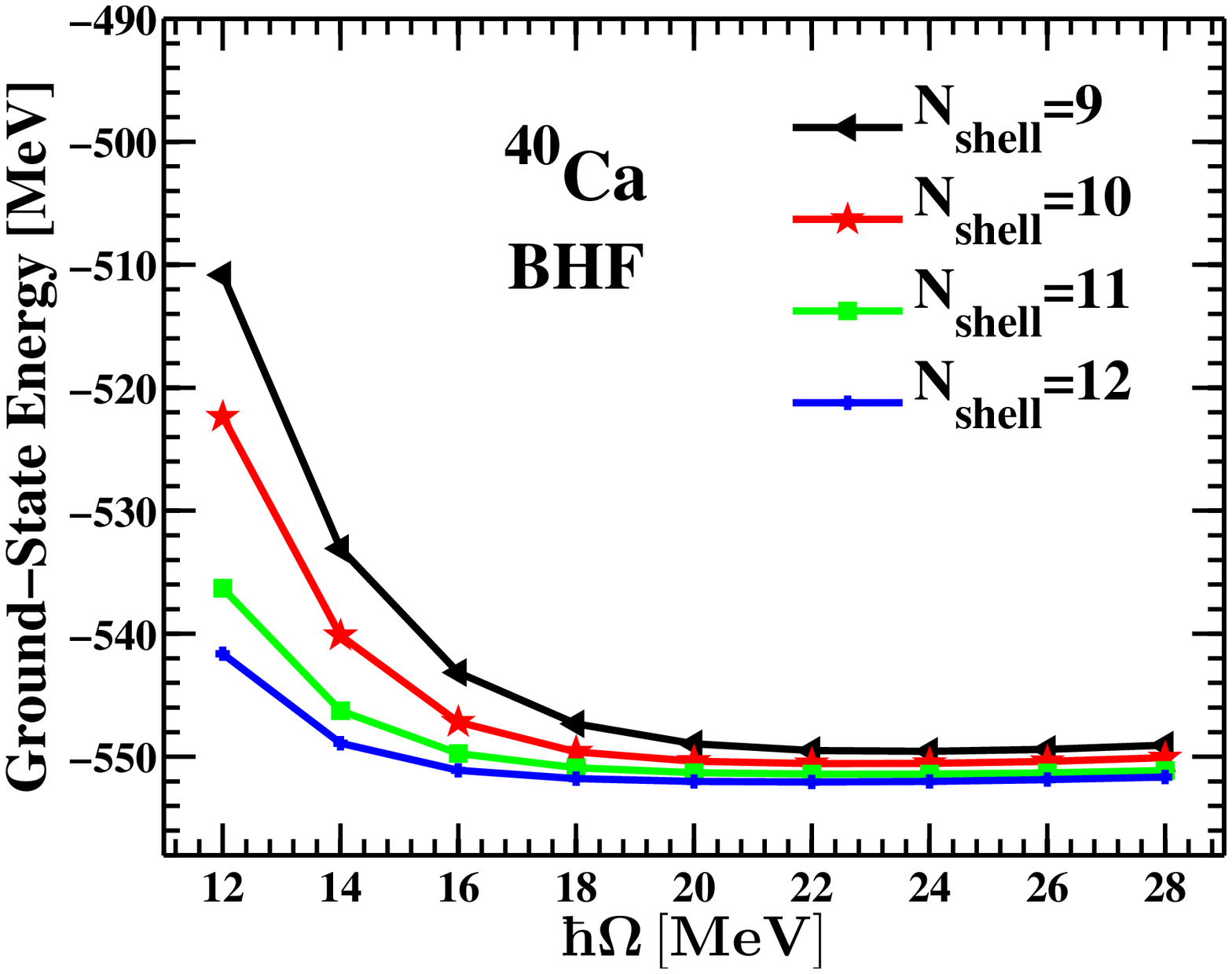}
\includegraphics[scale=0.43]{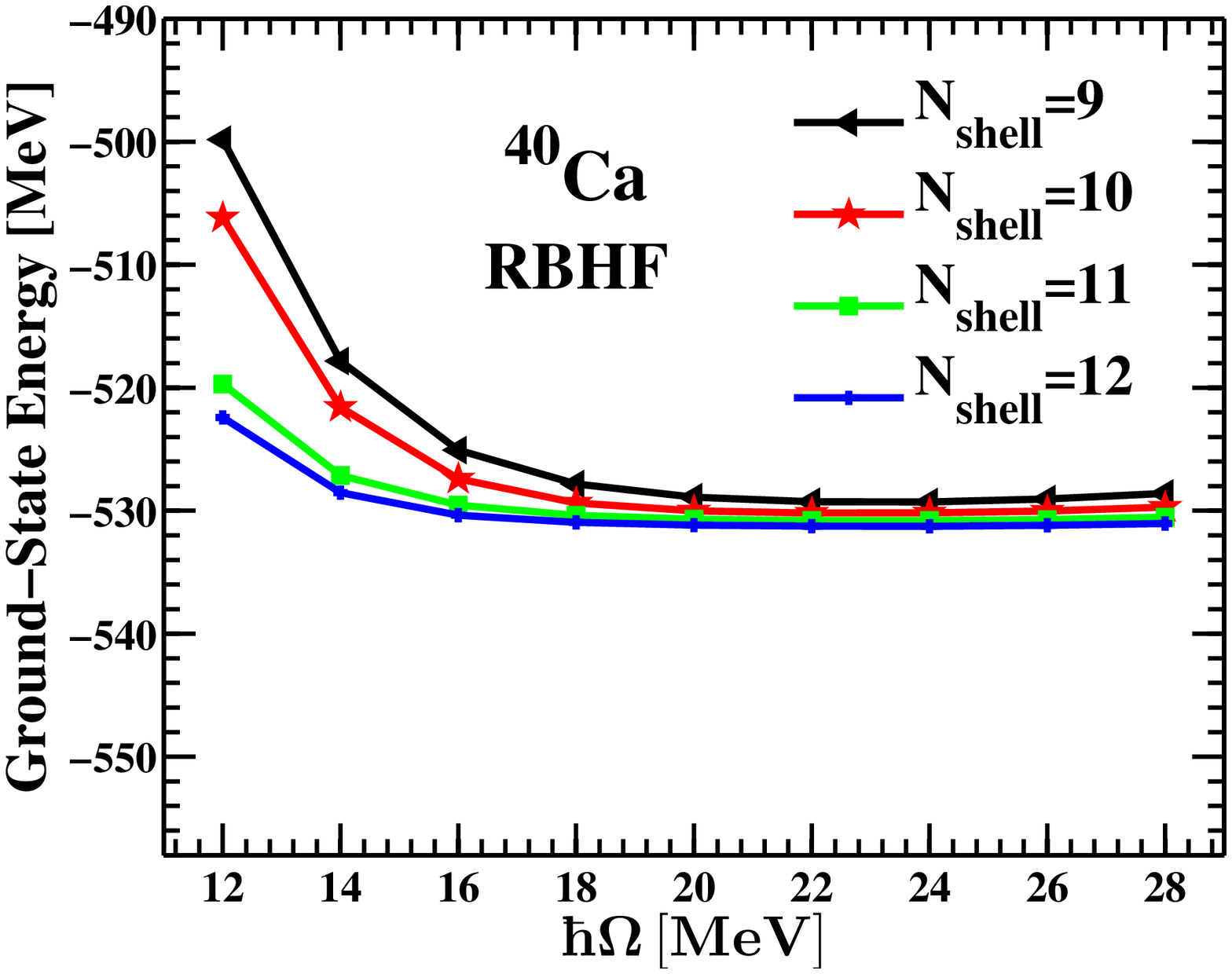}
\caption{\label{fig:bhf_rbhf} BHF and RBHF calculations of $^{4}$He, $^{16}$O and $^{40}$Ca
as a function of the oscillator parameter $\hbar \Omega$
with the  $V_{\text{low-}k}$ effective interaction derived from the Argonne $\upsilon_{18}$ \cite{PhysRevC.51.38} potential
at a sharp cutoff momentum $\Lambda$=1.9 fm$^{-1}$ for $^{4}$He and $\Lambda$=2.1 fm$^{-1}$ for $^{16}$O and $^{40}$Ca,
respectively.}
\end{figure}
Figure~\ref{fig:bhf_rbhf} shows the ground-state energies of $^{4}$He, $^{16}$O and $^{40}$Ca calculated  by BHF and RBHF
as a function of the oscillator parameter $\hbar \Omega$ with different
model space size, i.e., $N_{\text{shell}}$.
The BHF single-particle potential $U$ is taken by Eq.~(\ref{bhf potential}).
The elements of $G$-matrix, Eq.~(\ref{g-matrix}), are solved by matrix inversion method.
We see that the results of both BHF and RBHF by virtue of independent of the oscillator parameter $\hbar \Omega$
over a wide range (e.g., $\hbar \Omega \geq $18 MeV) are obtained for the different truncations $N_{\text{shell}}$.
For $N_{\text{shell}}$=11 and $\hbar \Omega$=22 MeV, the ground-state energy of $^{40}$Ca in BHF is -551.51 MeV.
While yielding a ground-state energy of -552.14 MeV with $N_{\text{shell}}$=12 at $\hbar \Omega$=22 MeV.
This shows that our results are converged within 0.5-1.0 MeV respected to the size of the model space.
The $N_{\text{shell}}$=12 calculations for these three nuclei appear nearly convergent.
The ground-state energies of RBHF
are higher than the results of BHF for all calculated nuclei.
This trend is opposite compared with most of the past RBHF calculations \cite{PhysRevC.4.81,PhysRevC.1.1644,PhysRevC.10.2080,PhysRevC.92.034312}.
We will see that if we take different description for the BHF single-particle potential $U$,
the self-consistent RBHF ground-state energy can also lower than BHF.
Our results can be understood from Eq.~(\ref{e_bhf}) and the alternative form of Eq.~(\ref{energy}),
\begin{eqnarray}
\label{e-bhf}
\begin{array}{ll}
  E_{\text{BHF}}=\displaystyle\sum_{h_{1}=1}^{A} \varepsilon_{h_{1}}
  -\dfrac{1}{2} \displaystyle\sum_{h_{1},h_{2}=1}^{A}
  \langle h_{1}h_{2} |G(\varepsilon_{h_{1}}+\varepsilon_{h_{2}})|h_{1}h_{2}\rangle.
  \end{array}
  \end{eqnarray}
\begin{eqnarray}
\label{e-rbhf}
E_{\text{RBHF}}=&&\displaystyle\sum_{h_{1}} \epsilon_{h_{1}} -
\frac{1}{2} \displaystyle\sum_{h_{1},h_{2}}
\langle h_{1}h_{2}|G(\epsilon_{h_{1}}+\epsilon_{h_{2}})|h_{1} h_{2}\rangle
P_{h_{1}}P_{h_{2}}.
\end{eqnarray}
The RBHF gives less attractive single-particle energies
yield larger summation in the first term on the right-hand side of the above equation~(\ref{e-rbhf}).
The main difference of the second term in Eq.~(\ref{e-rbhf}) comparing BHF, Eq.~(\ref{e-bhf}), is
the introduced self-consistent occupation probability $P_{h}$ 
which can suppress the change of $G$-matrix and give larger summation of the second term in Eq.~(\ref{e-rbhf}).
So the RBHF ground-state energy higher or lower than BHF
finally depends on the  difference value between the two terms in Eq.~(\ref{e-rbhf}).
Them depends on the description of the BHF single-particle potential $U$.

\begin{table}
\caption{
\label{comparison}
Binding energies (in MeV) for $^{4}$He, $^{16}$O and $^{40}$Ca calculated by different \emph{ab initio} methods
with the  $V_{\text{low-}k}$ effective interaction derived from the Argonne $\upsilon_{18}$ \cite{PhysRevC.51.38} potential.
We use a sharp cutoff $\Lambda$=1.9 fm$^{-1}$ for $^{4}$He and $\Lambda=2.1$ fm$^{-1}$ for $^{16}$O and $^{40}$Ca calculations,
respectively.
CC \cite{PhysRevC.76.044305}, HF-MBPT(3), BHF and RBHF calculations take
$N_{\text{shell}}$=12 and
the oscillator parameter $\hbar \Omega$ = 14 MeV in the case of $^{4}$He
and $\hbar \Omega$ = 22 MeV for $^{16}$O and $^{40}$Ca.
In IT-NCSM \cite{PhysRevLett.99.092501} calculations,
$\hbar \Omega$ = 22 MeV for $^{16}$O and $\hbar \Omega$ = 24 MeV for $^{40}$Ca
are taken.
}
\begin{ruledtabular}
\begin{tabular}{cccccccc}
\textrm{}&
\textrm{$^{4}$He}&
\textrm{$^{16}$O}&
\textrm{$^{40}$Ca}\\
\hline
 Experiment \cite{1674-1137-36-12-001} &  -28.30& -127.62& -342.05\\
 Exact (FY \cite{PhysRevC.70.061002,PhysRevC.76.044305}) &  -29.19(5)&$-$ & $-$\\
 IT-NCSM & $-$ & -138.0 & -462.7 \\
 CCSD  & -28.9 & -142.8 & -491.2\\
 CCSD(T) & -29.2 & -148.2 & -502.9\\
 HF-MBPT(3) & -29.33 & -159.34 & -600.08 \\
 BHF  & -25.90& -134.16& -552.14\\
 RBHF & -25.79& -130.04& -530.68\\
\end{tabular}
\end{ruledtabular}
\end{table}
Table~\ref{comparison} gives the quantitative comparison with the benchmark given by different {\it ab initio} calculations.
The calculations of FY, IT-NCSM, CCSD, CCSD(T), HF-MBPT(3), BHF and RBHF are included in this table.
HF-MBPT(3) labels the same calculation as shown in Ref.~\cite{PhysRevC.94.014303},
namely the MBPT corrections are up to third order in energy within the HF basis.
Nogga {\it et al.}~\cite{PhysRevC.70.061002}, first calculated the $^{4}$He ground-state energy
by solving the FY equations with only the two-body  $V_{\text{low-}k}$.
They estimated an accuracy of 50 keV for the $^{4}$He calculations.
So the results of FY can be regarded as an exact solution of  $^{4}$He to benchmark with other methods.
Roth and Navr$\acute{a}$til gave the IT-NCSM results of $^{16}$O and $^{40}$Ca in Ref. \cite{PhysRevLett.99.092501}.
In the IT-NCSM calculation, the model space truncation parameter $N_{\text{max}}$ was taken,
which measures the maximal allowed HO excitation energy above the unperturbed lowest zero-order reference
state.
They obtained a ground-state energy of -137.7 MeV
and a point-nucleon root-mean-square (rms) radius of 2.03 fm for $^{16}$O with $N_{\text{max}}$=14 and $\hbar \Omega$=22 MeV.
An exponential extrapolation of energy at this oscillator parameter yielded $E_{\infty}$=-138.0	MeV.
For $^{40}$Ca with $N_{\text{max}}$=16 and $\hbar \Omega$=24 MeV,
a ground-state energy of -461.8 MeV and a point-nucleon rms radius of 2.27 fm were obtained.
An exponential extrapolation yielded $E_{\infty}$=-462.7 MeV.
The CCSD and CCSD(T) ground-state energies in Table~\ref{comparison} are
the extrapolated infinite model space results \cite{PhysRevC.76.044305}.
In the HF-MBPT(3), BHF and RBHF calculations,
the basis spaces employed take $N_{\text{shell}}$=12
at $\hbar \Omega$ = 14 MeV in the case of $^{4}$He
and at $\hbar \Omega$ = 22 MeV for $^{16}$O and $^{40}$Ca.

From Table~\ref{comparison} we find that the ground-state energies of $^{4}$He calculated by BHF and RBHF
are similar, but higher than other calculations.
The results of IT-NCSM, CC and HF-MBPT are very closed to the exact FY calculation,
in good agreement with data.
For  $^{4}$He the number of the occupied single-particle states are few,
so the renormalized effects from the depletions of
the normally occupied single-particle states are small for ground-state energy.
For $^{16}$O and $^{40}$Ca the ground-state energies of HF-MBPT(3) are lower than the CC.
MBPT mostly offers a finite-order approximation to the many-body problems,
while CC theory provides an infinite-order approximation in selected cluster operators,
offering a very powerful re-summation of MBPT diagrams \cite{bartlett2009}.
The CCSD and CCSD(T) take about more correlations compare to MBPT(3).
Thus the higher order corrections of MBPT will draw back the ground-state energy.
The ground-state energies of BHF and RBHF are less bound compared with HF-MBPT(3).
In HF-MBPT(3) the third-order corrections are small comparing the second-order correction.
For example, in $^{40}$Ca calculation,
-110.09 MeV for second-order correction,
-5.37 MeV for hole-hole diagram, (c) of Fig.~1 in Ref.~\cite{PhysRevC.94.014303},
and 12.35 MeV for particle-hole diagram, (e) of Fig.~1 in Ref.~\cite{PhysRevC.94.014303}.
The correction of ladder diagram in third order are very small, e.g., -1.24 MeV for $^{40}$Ca.
Thus the two diagrams, hole-hole and particle-hole diagrams,
which not included in the BHF (RBHF) have small contribution to the total energy.
Ignoring the hole-hole and particle-hole diagrams in third-order corrections of MBPT,
the BHF and RBHF take higher order corrections than  HF-MBPT(3), yielding  less ground-state energy.


\begin{table}[h]
\caption{
\label{rms}
Point-nucleon rms radii (in fm) of doubly magic nuclei in different calculations.
The results of IT-NCSM are taken from Ref. \cite{PhysRevLett.99.092501}.
The effective interaction and oscillator parameter $\hbar \Omega$ are same as Table~~\ref{comparison}.
}
\begin{ruledtabular}
\begin{tabular}{cccccccc}
\textrm{Nucleus}&
\textrm{IT-NCSM}&
\textrm{BHF}&
\textrm{RBHF}\\
\hline
 $^{4}$He  &  $-$ & 1.22 & 1.28\\
 $^{16}$O  &  2.03& 1.92 & 2.05\\
 $^{40}$Ca &  2.27& 2.20 & 2.30\\
\end{tabular}
\end{ruledtabular}
\end{table}
Table~\ref{rms} gives the IT-NCSM, BHF and RBHF calculations for
the point-nucleon rms radii of $^{4}$He, $^{16}$O and $^{40}$Ca.
In BHF and RBHF, a simple and frequently used center-of-mass correction method is used to
remove the center-of-mass motion component of the rms radius as shown in Ref.~\cite{PhysRevC.69.034332}.
We see that the radii of RBHF are larger than BHF results for all calculated nuclei,
and the RBHF has the similar radii with the IT-NCSM.
As the description in Introduction,
the RBHF approach takes into account the depletions of
the normally occupied single-particle states
resulting from many-body correlations.
When the occupation probabilities $P_{h}$, Eq.~(\ref{p}), are taken into the BHF single-particle potential $U$,
the single-particle potential will be smaller attractive.
Then the iterative solution of BHF equations~(\ref{bhf_eqs}) will give
more occupied single-particle energies, less kinetic energy and
less gap between occupied and unoccupied states, comparing the RBHF to BHF.
The radius of the nucleus becomes larger when the kinetic energy is reduced.

\begin{table}
\caption{
\label{spe_o16}
Single-particle energies $\varepsilon$ (in MeV) and occupation probabilities $P$ for $^{16}$O.
The experimental data are taken from Refs. \cite{nndc,PhysRevLett.33.431}.
}
\begin{ruledtabular}
\begin{tabular}{cccccccc p{cm}}
\multicolumn{2}{c}{} & \multicolumn{6}{c}{$^{16}$O} \\
\multicolumn{2}{c}{} & \multicolumn{3}{c}{Neutron} & \multicolumn{3}{c}{Proton} \\
 Orbital & & BHF & RBHF & Expt.  & BHF & RBHF & Expt. \\
 \hline
  $1 s_{1/2}$   &$ \varepsilon$&-73.02&-59.04& -47    &-72.75&-58.84& $-44\pm 7$\\
                &$P$          &  1.00&  0.93&        &  1.00&  0.93&           \\
  ${1 p_{3/2}}$ &$ \varepsilon$&-37.38&-27.85& -21.839&-37.15&-27.67&-18.451    \\
                &$P$          &  1.00&  0.87&        &  1.00& 0.87&           \\
  ${1 p_{1/2}}$ &$ \varepsilon$&-24.35&-18.50& -15.663&-24.16&-18.35&-12.127    \\
                &$P$          &  1.00&  0.87&        &  1.00& 0.87&           \\
  ${1 d_{5/2}}$ &$\varepsilon$& -4.65& -0.34& -4.144 & -4.49&-0.23&-0.601     \\
                &$P$          &  1.00&  1.00&        &  1.00& 1.00&           \\
  ${2 s_{1/2}}$ &$\varepsilon$& -2.46& -0.47& -3.273 & -2.34&-0.39&-0.106     \\
                &$P$          &  1.00&  1.00&        &  1.00& 1.00&           \\
  ${1 d_{3/2}}$ &$\varepsilon$&  6.84&  7.28& 0.941  &  6.90& 7.33&4.399      \\
                &$P$          &  1.00&  1.00&        &  1.00& 1.00&           \\
\end{tabular}
\end{ruledtabular}
\end{table}
\begin{table}
\caption{
\label{spe_ca40}
Single-particle energies $\varepsilon$ (in MeV) and occupation probabilities $P$ for $^{40}$Ca.
The experimental data are taken from Refs. \cite{nndc,PhysRevLett.33.431}.
}
\begin{ruledtabular}
\begin{tabular}{cccccccc p{cm}}
\multicolumn{2}{c}{} & \multicolumn{6}{c}{$^{40}$Ca} \\
\multicolumn{2}{c}{} & \multicolumn{3}{c}{Neutron} & \multicolumn{3}{c}{Proton} \\
 Orbital & & BHF & RBHF & Expt.  & BHF & RBHF & Expt. \\
 \hline
  ${1 s_{1/2}}$ &$\varepsilon$&-137.71&-116.61& $-$    &-137.14&-116.15&$-49.1\pm 12$ \\
                &             &      &      &        &      &      &$-77\pm 14$   \\
                &$P$          &  1.00&  0.97&        &  1.00&  0.97 &            \\
  ${1 p_{3/2}}$ &$\varepsilon$&-92.34&-74.75& $-$    &-91.84& -74.35&$-33.3\pm 6.5$    \\
                &$P$          &  1.00&  0.93&        &  1.00&  0.93&            \\
  ${1 p_{1/2}}$ &$\varepsilon$&-74.62&-60.95& $-$    &-74.16& -60.59& $-32\pm 4$ \\
                &$P$          &  1.00&  0.93&        &  1.00&  0.93&            \\
  ${1 d_{5/2}}$ &$\varepsilon$&-53.13& -39.32& -21.30 &-52.70& -39.00&$-14.9\pm 2.5$   \\
                &             &      &      &        &      &      &$-13.8\pm 7.5$  \\
                &$P$          &  1.00&  0.85&        &  1.00&  0.85&            \\
  ${2 s_{1/2}}$ &$\varepsilon$&-43.33& -32.75& -18.104&-42.88& -32.41&-10.850     \\
                &$P$          &  1.00&  0.87&        &  1.00&  0.87&            \\
  ${1 d_{3/2}}$ &$\varepsilon$&-28.15& -20.47& -15.635&-27.80& -20.20&-8.328      \\
                &$P$          &  1.00&  0.79&        &  1.00&  0.79&            \\
  ${1 f_{7/2}}$ &$\varepsilon$&-16.21& -7.26& -8.363 &-15.88& -7.01&-1.085      \\
                &$P$          &  1.00&  1.00&        &  1.00&  1.00&            \\
  ${2 p_{3/2}}$ &$\varepsilon$&-10.45&  -5.29& -6.420 &-10.16&-5.09&0.631       \\
                &$P$          &  1.00&  1.00&        &  1.00&  1.00&            \\
  ${2 p_{1/2}}$ &$\varepsilon$& -3.06&  -0.08& $-$    &-2.85& 0.07&$-$         \\
                &$P$          &  1.00&  1.00&        &  1.00&  1.00&            \\
  ${1 f_{5/2}}$ &$\varepsilon$&  8.26&   9.66& $-$    & 8.40& 9.75&$-$         \\
                &$P$          &  1.00&  1.00&        &  1.00&  1.00&            \\
\end{tabular}
\end{ruledtabular}
\end{table}
Becker and Patterson \cite{Becker197188} have pointed out that
the near equality of single-particle energies and separation energies which holds in RBHF
(as in Koopmans' theorem \cite{KOOPMANS1934104}) but fails badly in BHF.
Thus for RBHF the single-particle energies can be directly related to the experimental mean removal energies
\cite{Wagner1973,doi:10.1146/annurev.ns.25.120175.000245}.
The experimental single-particle energies are observed
in knockout, stripping and pickup reactions,
primarily for these states close to the Fermi level.
The single-particle energies and occupation probabilities of BHF and RBHF calculations for $^{16}$O and $^{40}$Ca
are shown in Tables~\ref{spe_o16} and ~\ref{spe_ca40}, respectively.
In the RBHF, less bound single-particle energies with little change in ground-state energy  are got.
We can find renormalization with occupation probabilities has a dramatic effect upon the single-particle levels.
It can lead to the level density increases and the levels higher.
Both calculations give single-particle levels of $^{16}$O and $^{40}$Ca
which are more bound than the experimental data.
We use the $NN$-only $V_{\text{low-}k}$ interaction,
omitting the three-body and higher-order forces.
So the ground-state energy is over-binding, compared with data.
Thus the more bound single-particle energies are reasonable.

\begin{table}
\caption{
\label{cases}
Binding energies (in MeV) obtained by the BHF and RBHF with different prescriptions.
Three cases for the definition of BHF single-particle potential $U$ and the solving method of Bethe-Goldstone Eq.~(\ref{g})
are chosen.
The effective interaction, $N_{\text{shell}}$ and $\hbar \Omega$ are same as Table~~\ref{comparison}.
}
\begin{ruledtabular}
\begin{tabular}{ccccc p{cm}}
  Nucleus & Method & Case (1) & Case (2)& Case (3)\\
\colrule
  \multirow{2}{*}{$^{4}$He}
  &BHF & -25.90&  -25.83& -25.06\\
  &RBHF& -25.79&  -25.72& -25.29\\
  \hline
  \multirow{2}{*}{$^{16}$O}
  &BHF & -134.16&  -134.07& -125.39\\
  &RBHF& -130.04&  -129.97& -127.83\\
    \hline
  \multirow{2}{*}{$^{40}$Ca}
  &BHF & -552.14&  -552.16& -529.02\\
  &RBHF& -530.68&  -530.83& -533.85\\
\end{tabular}
\end{ruledtabular}
\end{table}
We also perform BHF and RBHF by using different prescriptions:

Case (1) The BHF single-particle potential $U$ is taken by Eq.~(\ref{bhf potential}).
The elements of $G$-matrix, Eq.~(\ref{g-matrix}), are solved by inversion method.

Case (2) The BHF single-particle potential $U$ is taken by Eq.~(\ref{bhf potential}).
The elements of $G$-matrix, Eq.~(\ref{g-matrix}), are solved by iteration.
In this iteration method,
the $G$-matrix, Eq.~(\ref{g}), is expressed as a sum of terms
\begin{eqnarray}
\label{iteration}
G(\omega)=\hat{V}+\hat{V} \dfrac{Q}{e} \hat{V} + \hat{V} \dfrac{Q}{e} \hat{V}\dfrac{Q}{e} \hat{V} + \ldots.
\end{eqnarray}
We express this sum up to the third terms in this work.

Case (3)
The BHF single-particle potential $U$ is taken as the following form,
\begin{eqnarray}
\langle a|U|b \rangle=
\left\{
\begin{array}{lll}
\dfrac{1}{2} \displaystyle\sum_{h \leq \varepsilon_{F}}
\langle ah| G(\varepsilon_{a}+\varepsilon_{h})+G(\varepsilon_{b}+\varepsilon_{h})
|bh\rangle \qquad \text{for} \; a,b\leq \varepsilon_{F}\\
\displaystyle\sum_{h \leq \varepsilon_{F}}\langle ah| G(\varepsilon_{a}+\varepsilon_{h}))
|bh\rangle  \qquad \; \quad \quad \qquad \qquad \text{for} \; a  \leq \varepsilon_{F}, b > \varepsilon_{F}\\
\displaystyle\sum_{h \leq \varepsilon_{F}}\langle ah| G(\varepsilon_{b}+\varepsilon_{h}))
|bh\rangle  \qquad \; \quad \quad \qquad \qquad \text{for} \; a  > \varepsilon_{F}, b \leq \varepsilon_{F}\\
0 \qquad \qquad \qquad \qquad \qquad \qquad \qquad \qquad \qquad \text{for} \; a,b > \varepsilon_{F}
\end{array} \right. .
\end{eqnarray}
The elements of $G$-matrix, Eq.~(\ref{g-matrix}), are solved by inversion method.

The comparison among the above three cases is listed in the Table~\ref{cases}.
Case (1) and case (2)  give almost the same results.
The results,
especially for  $^{16}$O and $^{40}$Ca,
have significant differences for different definition of BHF single-particle potential $U$ in BHF calculations.
While the RBHF calculations give small differences for all calculated nuclei.
This implies the self-consistent occupation probabilities can suppress
the effect from change of $G$-matrix when choosing different particle-particle elements of $U$.
We can conclude that, on the one hand, the iterative solution of the Bethe-Goldstone Eq.~(\ref{g}) converges very quickly,
and the inversion method to solve Bethe-Goldstone Eq.~(\ref{g}) is stable.
On the other hand, as long as the particle-particle elements $\langle p_{1}|U| p_{2} \rangle$ is small,
the RBHF results have little influence from the different definitions of $\langle p_{1}|U| p_{2} \rangle$.

\section{\label{sec:summary}Summary}

We have performed the self-consistent Brueckner-Hartree-Fock (BHF) and Renormalized Brueckner-Hartree-Fock (RBHF) calculations for finite nuclei with realistic $NN$ interaction.
``Self-consistent'' implies that we calculate the $G$-matrix within the BHF (or RBHF) basis,
and the Pauli exclusion operator is determined by the BHF (or RBHF) spectrum.
The RBHF by renormalization with occupation probabilities resulting from many-body correlations
is especial for single-particle energies and radius.
The $G$-matrix is calculated using the $V_{\text{low-}k}$ effective interaction
derived from Argonne $\upsilon_{18}$ potential \cite{PhysRevC.51.38}.
Different techniques are taken to solve the Bethe-Goldstone equation.
We conclude that the iterative solution of the Bethe-Goldstone equation has very fast convergence  for soft force,
and the inversion method is stable.
The particle-particle matrix elements of BHF potential energy are a somewhat controversial matter,
because the corresponding diagrams require an off-shell definition of the starting energy.
In the past experience, if the particle-particle matrix elements are not very large,
three-body correlations can be effectively summed.
Different prescriptions for particle-particle matrix elements of BHF potential
have been calculated in this work.
We find that if the particle-particle matrix elements are small,
even set to zero, their affection is small for the results.
Based on past calculations, the RBHF give more bound energies and larger radius than BHF results.
However, we find that different descriptions to the particle-particle matrix elements of BHF potential energy
can give more or less bound energies, while the radius is consistently larger.

We first give the benchmark calculation for BHF and RBHF with other \emph{ab initio} methods.
The closed-shell nuclei $^{4}$He, $^{16}$O and $^{40}$Ca have been chosen as examples for the present calculations.
The solution of BHF is a complicated doubly self-consistent procedure:
(i) Calculate the $G$-matrix via Eq.~(\ref{g-matrix}) in a suitable basis of first choice (e.g., Woods-Saxon basis in this work);
(ii) Solve the BHF equations~(\ref{bhf_eqs}), which give a new basis;
(iii) Calculate a new $G$-matrix in this new basis;
and so on until the convergence is achieved.
In RBHF, three self-consistent should be satisfied,
\emph{i.e.}, $G$-matrix self-consistent, HF self-consistent and occupation probability self-consistent
are coupled together.
The convergences with respect to the HO frequency and model truncation have been discussed in details.
The general results are consistent with other \emph{ab initio} methods,
e.g., Faddeev-Yakubovsky equations, no-core shell model and coupled cluster.
Our results confirm that the BHF and RBHF are a powerful method to derive the bulk properties of nuclear systems.
Renormalization with occupation probabilities is crucial for getting a reasonable single-particle spectrum.
However, we use $NN$-only force in this work.
Three-body and higher-order forces are not considered.
So we get the over-bound ground-state energies and single-particle energies of $^{16}$O and $^{40}$Ca.
The questions worth examining in the future are that
making a similar comparison between self-consistent BHF and other  \emph{ab initio} method for heavier nuclei
with three-body Hamiltonians including full three-body forces.

\begin{acknowledgments}
This work has been supported by
the National Key Basic Research Program of China under Grant No. 2013CB834402;
the National Natural Science Foundation of China under Grants No. 11235001, No. 11320101004 and NO. 11575007;
the CUSTIPEN (China-U.S. Theory Institute for Physics with Exotic Nuclei) funded by the U.S.  Department of Energy, 
Office of Science under grant number DE-SC0009971.
\end{acknowledgments}

\bibliography{references}
\end{document}